\documentclass[%
 superscriptaddress,
 reprint,
 linenumbers,
 showpacs,preprintnumbers,
 nofootinbib,
 amsmath,amssymb,
 aps,
 prd,
 floatfix,
]{revtex4-1}
\usepackage{graphicx}
\usepackage{xcolor}
\usepackage{multirow}
\usepackage{subfig}
\usepackage{amssymb,amsmath,amstext,amsthm,amsfonts}
\usepackage[utf8]{inputenc}
\usepackage{float}
\usepackage{url}
\usepackage{soul}

\usepackage{todonotes}

\definecolor{LightCyan}{rgb}{0.88,1,1}
\usepackage{xcolor,colortbl}
\usepackage{placeins}

\begin{document}
\nolinenumbers
	\title{Adversarial methods to reduce simulation bias in neutrino interaction event filtering at Liquid Argon Time Projection Chambers}

	\author{M. Babicz}
	\email[Contact e-mail: ]{marta.babicz@cern.ch.}
	\affiliation{H. Niewodniczanski Institute of Nuclear Physics PAN, ul. Radzikowskiego 152, Krakow, Poland}
    \affiliation{CERN, Espl. des Particules 1, Meyrin, Switzerland}
    
	\author{S. Alonso-Monsalve}
	\email[Contact e-mail: ]{saul.alonso.monsalve@cern.ch. The author's work presented in the manuscript was done when the author was a member of the CERN neutrino group.}
	\affiliation{ETH Zurich, Rämistrasse 101, Zürich, Switzerland}
	\affiliation{CERN, Espl. des Particules 1, Meyrin, Switzerland}
	
	\author{S. Dolan}
	\email[Contact e-mail: ]{stephen.joseph.dolan@cern.ch.}
	\affiliation{CERN, Espl. des Particules 1, Meyrin, Switzerland}
	
	\author{K. Terao}
	\email[Contact e-mail: ]{kterao@slac.stanford.edu.}
	\affiliation{SLAC National Accelerator Laboratory, Menlo Park, CA, 94025, USA}

\begin{abstract}
\noindent For current and future neutrino oscillation experiments using large Liquid Argon Time Projection Chambers (LAr-TPCs), a key challenge is identifying neutrino interactions from the pervading cosmic-ray background. 
Rejection of such background is often possible using traditional cut-based selections, but this typically requires the prior use of computationally expensive reconstruction algorithms. This work demonstrates an alternative approach of using a 3D Submanifold Sparse Convolutional Network trained on low-level information from the scintillation light signal of interactions inside LAr-TPCs. This technique is applied to example simulations from ICARUS, the far detector of the Short Baseline Neutrino (SBN) program at Fermilab. The results of the network, show that cosmic background is reduced by up to 76.3\% whilst neutrino interaction selection efficiency remains over 98.9\%. 
We further present a way to mitigate potential biases from imperfect input simulations by applying Domain Adversarial Neural Networks (DANNs), for which modified simulated samples are introduced to imitate real data and a small portion of them are used for adverserial training. A series of mock-data studies are performed and demonstrate the effectiveness of using DANNs to mitigate biases, showing neutrino interaction selection efficiency performances significantly better than that achieved without the adversarial training.

\end{abstract}

\maketitle

\section{Introduction}
\vspace{-0.1cm}
\label{sec:intro}
The current and next generation of neutrino oscillation experiments offer a tantalising opportunity to explore physics beyond the Standard Model. However, as detectors grow larger and neutrino beams more powerful, a pre-filtering of relevant neutrino interaction data becomes increasingly important for experiments to be computationally viable. This is particularly crucial for liquid Argon time projection chambers (LAr-TPCs), which are commonly used neutrino detectors (see e.g.,~\cite{Rubbia:1977zz, Willis:1974gi, ArgoneuT:2016wjb, Antonello:2015lea, Abi:2020wmh}). 
LAr-TPCs offer precise spatial and calorimetric measurements based on the electron drift signal from ionisation and the scintillation photons from the excitation of Argon atoms caused by interacting particles. However, detecting neutrino interactions with this technology becomes challenging due to the significant background of incoming cosmic rays.

For example, at the ICARUS detector of the SBN experiment, cosmic rays are expected to outnumber neutrino interactions within the Booster Neutrino Beam's (BNB) spill gate by more than three to one~\cite{Antonello:2015lea}. Even located 1.5 km underground, the DUNE far detectors will experience a comparable rate of cosmic rays and neutrinos~\cite{Abi:2020wmh}.  

In LAr-TPCs, the ionisation electrons, stimulated by propagating charged particles, are drifted by an applied electric field to be collected by TPC anode wires, whilst the emitted LAr scintillation light is recorded by photodetectors, often using Photomultiplier Tubes (PMT). As such, the TPC records charged particle trajectories as images with a high spatial resolution ($\sim$mm/pixel) and the photodetectors provide event timing information with nanosecond resolution. The scintillation light signal thereby provides an easily accessible means to classify events requiring little or no processing, which may help distinguish cosmic rays from neutrino interactions before running any reconstruction algorithms.

The rejection of cosmic-ray backgrounds in LAr-TPCs typically starts at the online stage. A trigger to record data is issued only if a fast signal from the photodetectors is observed in the beam spill window, the time window during which neutrino signal is expected. However, it does not prevent the selection of background events caused by an accidental coincidence of a beam spill window with incident cosmic rays.
More sophisticated filtering may be achieved through multi-dimensional analyses, as discrimination power can be found by analysing which PMTs received light, at what time, and for how long relative to all other PMTs inside the LAr-TPC. 

Machine-learning methods, capable of analysing such high dimensional information, are therefore excellently suited to classifying events using the available PMT information. In particular, when detector data is represented as images, the use of Convolutional Neural Networks (CNNs)~\cite{firstCNN,LeCun-et-al-1998-gradient} are especially effective. The use of CNNs for event classification is well established across the field of neutrino physics~\cite{Aurisano:2016jvx, DUNE:2020gpm, MicroBooNE:2018kka}. Although CNNs are broadly used in the neutrino community, images of neutrino interactions are typically very sparse such that most of the pixels have empty values which can render some standard methods ineffective. A straightforward solution to this issue is to use submanifold sparse convolutional networks (SSCNs)~\citep{graham2017submanifold}, a variation of standard CNNs that use a new sparse convolutional operator to efficiently handle sparse inputs, with already a remarkable number of successful applications in neutrino physics~\cite{Adams-2019-deep,Domine-2020-scalable}. All the neural network architectures shown in this paper belong to the class of SSCNs. Nevertheless, other architectures, such as graph neural networks~\cite{SperdutiFirstGNN,zhou2018graph}, were also considered but were found to produce significantly worse results in comparison to SSCNs for our application (although further optimisation of these alternatives may have been able to improve performance).

CNNs\footnote{We refer to CNNs and SSCNs indistinguishably in the rest of this section.} can be optimized to discriminate signal images against backgrounds through a supervised training process. This is often done using simulated images (e.g., simulated neutrino and cosmic images) where the true labels are available. However, when this model is applied to images from the real detector, its performance is typically worse than what is observed on simulated images because of discrepancies between two data domains (i.e., physics of the real world v.s. simulation) due to imperfect simulation.
To address this problem, the CNN classifier may rely on domain adaptation (DA) techniques~\cite{ben2010theory, redko2019advances} so that the classifier learned from the training domain (i.e., simulated data) can also be applied to the testing domain (i.e., eventual experimental data). This DA can be achieved through the application of Domain Adversarial Neural Networks (DANNs)~\cite{dann}, in which the detector data is used in an unsupervised (or semi-supervised) manner to prevent the CNN exploiting features that differ between data and simulation. DANNs were first used in neutrino physics by the MINER$\nu$A experiment, where the bias of a deep-learning-based neutrino vertex identification method was mitigated using these techniques~\cite{perdue2018reducing}. In this manuscript, we present the first application of DANN for a CNN as an event classifier for a LAr-TPC to discriminate neutrino signal against cosmic backgrounds.

To test the effectiveness of CNNs and DANNs at distinguishing cosmic-ray backgrounds from neutrino interactions using only information from the scintillation signal, we consider the ICARUS detector of the SBN program~\cite{Antonello:2015lea} as a case study. ICARUS is currently the world's largest LAr-TPC employed in neutrino physics and operates close to the surface, and so is subject to a particularly challenging cosmic-ray background rejection. The details of the ICARUS detector and simulation are summarised in Sec.~\ref{sec:icarus}. The CNN approach to event filtering is detailed and demonstrated in Sec~\ref{sec:cnnmethod}. The application of DANNs to reduce the CNN sensitivity to input simulation dependence, and a method of using \textit{mock-data} studies to test their effectiveness, is then described and applied in Sec~\ref{sec:dannmethod}. Finally, the results and the main conclusions of this work are presented in Sec~\ref{sec:discussion}.

\section{Event filtering at the ICARUS detector}
\label{sec:icarus}

\begin{figure}[htb]
    \centering
    \includegraphics[width=0.85\linewidth]{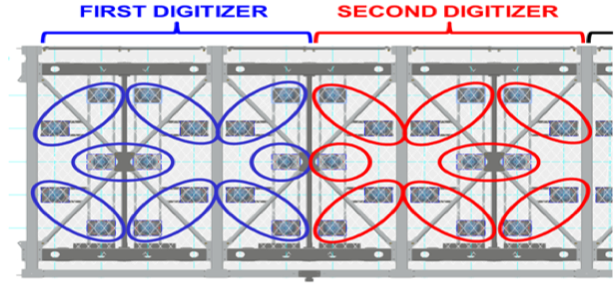}
    \caption{A schematic view of a one wall segment of ICARUS TPCs showing how the PMTs are distributed. The PMTs are arranged in groups of 15 that are connected to the same digitiser board. The hollow ellipses show the pairing system of adjacent PMTs.}
    \label{fig:PMT_pairing_scheme}
\end{figure}

The ICARUS detector~\cite{Amerio:2004ze} is a 760-ton LAr-TPC, serving as the far detector of the SBN program~\cite{Antonello:2015lea}, positioned 600 m away from the Booster Neutrino Beam (BNB) at FNAL. The detector consists of two identical adjacent modules, each housing two TPCs separated by a common cathode used to generate the electric field that directs the Argon ionisation signal to the anode. The prompt (order of nanosecond) LAr scintillation light signal from charged particles propagating within ICARUS is readout by 360 8 inches PMTs~\cite{Babicz:2018svg} arranged on the walls of the TPCs, placed as shown in Fig.~\ref{fig:PMT_pairing_scheme}. The PMT system provides the means to trigger the readout of signals within the 1.6 $\mu$s beam spill windows whilst also enabling fast spatial localisation of neutrino beam associated events. The placement, performance and timing resolution of the PMTs are expected to allow the localisation of the associated charged particle track with accuracy better than 1~m~\cite{Babicz:2018svg, Ali-Mohammadzadeh:2020fbd, Babicz:2020rpf}. The signals of all PMTs are continuously readout, where pairs of adjacent PMTs are typically used together within the ICARUS trigger system (the pairing scheme is shown in Fig.~\ref{fig:PMT_pairing_scheme}). For each beam spill window, the ICARUS trigger system can assess which PMTs have a signal exceeding a predefined threshold, at what time that signal was recorded with respect to the start of the beam window and how many times the PMT recorded an \textit{opening} above the threshold. Note that a new opening is counted every 0.16 $\mu$s that the PMT signal remains above a pre-set threshold and so the number of openings acts as a discretised measurement of the time over the threshold which itself is highly correlated with the signal amplitude.

\vspace{-0.1cm}

\begin{figure*}[htb]
\centering
\vspace{-0.4cm}
\includegraphics[width=0.8\linewidth]{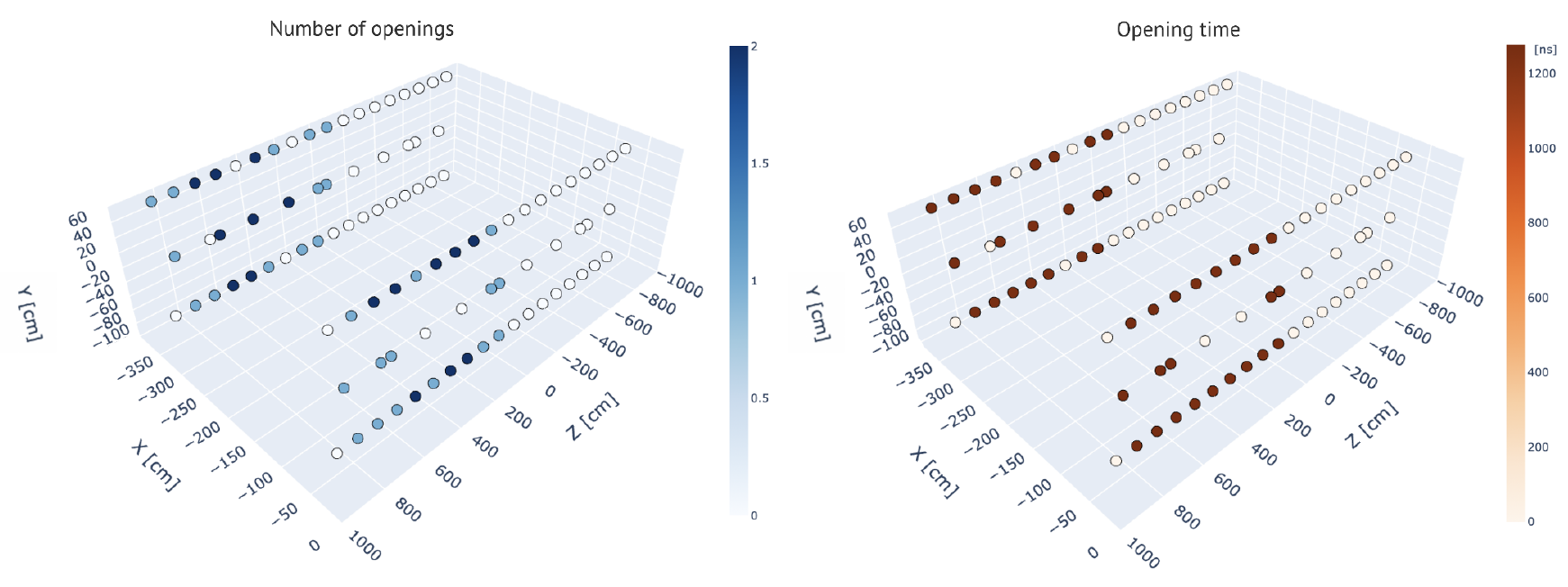}
\vspace{-0.15cm}
\caption{Example images of one ICARUS cryostat used as an input to the CNN. Each dot represents a PMT pair position (taken as the pair's barycentre) which are distributed across the walls of the TPCs. The colour of the dots represents the number of openings (left) or the opening time (right). }
\vspace{-0.1cm}
\label{fig:inputimages}
\end{figure*}

The BNB delivers neutrinos in a bunch at the rate up to 5 Hz. The accelerator complex informs the arrival time of each bunch, and hence the timing of the beam spill window, during which the detector may issue a trigger if there is a signal in photodetectors above the threshold. This results in about 5\% of beam spills being recorded. Despite a large reduction factor, the recorded events are still dominated by cosmic-ray backgrounds due to an accidental coincidence of the beam window and optical signals produced by cosmic rays. Our goal is to further reduce the cosmic-ray backgrounds via CNN-based fast event filter that only requires optical data.

Given the possibility of better separation of neutrino interactions from cosmic rays after full reconstruction of the TPC signal, the primary aim of the low-level event filter is to reduce the vast majority of cosmic rays while not compromising neutrino selection efficiency. In this way, the amount of data that needs to be processed for higher-level analyses is greatly reduced while avoiding the risk of losing the neutrino signal. In order to allow the possibility of \textit{online} event filtering, only the information available to the ICARUS event trigger is used. This means that information from each PMT pair is concatenated where the earliest opening time from each pair is stored alongside the total number of openings across both PMTs within the pair. 

\vspace{-0.5cm}

\subsection{Simulation}
\label{sec:simulation}

\vspace{-0.1cm}

The cosmic-ray particles impinging the ICARUS detector were generated with CORSIKA event generator~\cite{Heck:1998vt}. These particles are then propagated through the ICARUS detector and the surrounding material using GEANT4~\cite{Agostinelli:2002hh} implementation in LArSoft~\cite{Snider:2017wjd}. Scintillation photons are then propagated to the PMTs using a parameterised model based on pre-calculated tables (also derived from GEANT4). A further parameterised PMT readout model, constrained from test-beam data, is then used to simulate the digitised data-like signal from the detector.  

The incoming flux of neutrinos is modelled using a GEANT4-based simulation of the BNB beamline~\cite{Antonello:2015lea, AguilarArevalo:2008yp} whilst their interactions with nuclei (and electrons) within ICARUS are described using GENIE version~3~\cite{Andreopoulos:2009rq}. The particle propagation and detector response are simulated identically to the case of cosmic rays.

For this study, we simulate 396,200 PMT readout windows (\textit{events}) containing cosmic rays and 120,000 containing a single neutrino interaction. For this work, only one of ICARUS' two cryostats is considered. Whilst it is possible to have a PMT readout window containing cosmic rays and a neutrino interaction or multiple neutrino interactions, this is not particularly common, and such details are beyond the scope of this study. The number of events used in this work was reduced to those that passed the ICARUS trigger conditions, resulting in 114,589 neutrino and 46,115 cosmic events.

\section{CNN event filter}
\label{sec:cnnmethod}

The goal of the CNN is to classify whether events are from neutrino or cosmic-ray interactions. To train the CNN, the simulated PMT data is presented as 3D images, where the position of each PMT pair, alongside its opening time and a number of openings, are stored. The image pixel size is chosen to be 40~cm, which is the maximum distance such that two PMT pairs do not appear within the same pixel. An example of the image provided as input, divided into two sub-images representing the two PMT pair observable for better visualisation, is shown in Fig.~\ref{fig:inputimages}. Each event contains one image, which expresses both the opening time and number of openings for each PMT pair, of which 80\% are used for training, 10\% for validation, and the remaining 10\% for testing. Each image in the training sample is labelled as a cosmic-ray or neutrino event.

\begin{figure*}[htb]
\centering
\vspace{-0.3cm}
\includegraphics[width=0.93\linewidth]{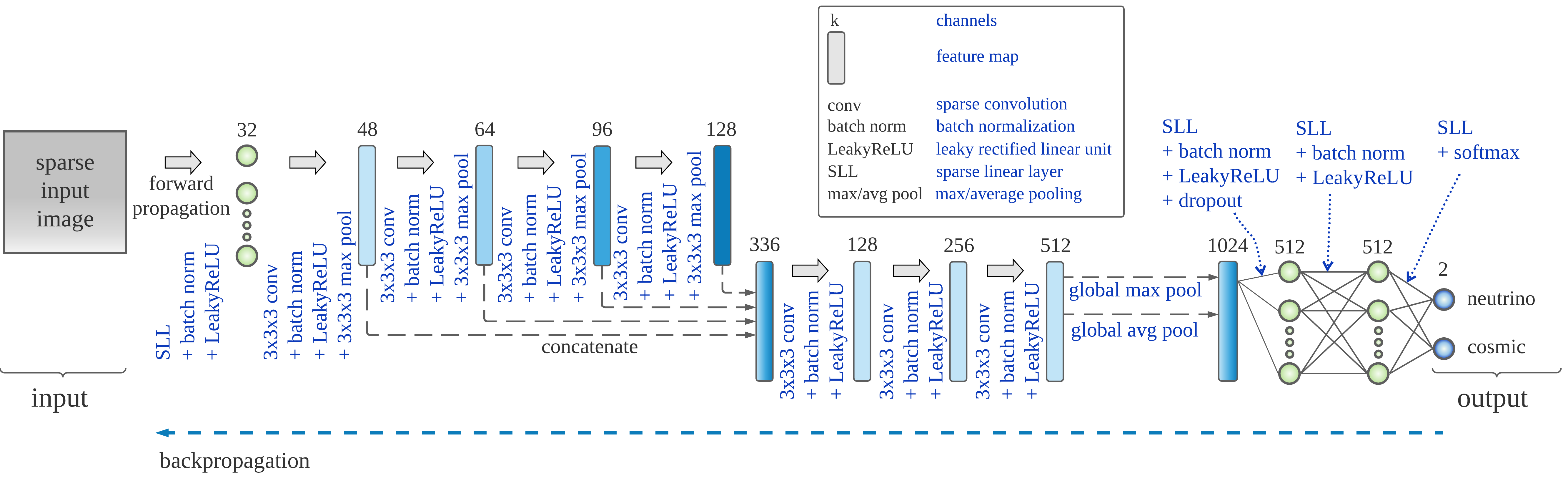}
\vspace{-0.2cm}
\caption{The sparse convolutional network architecture used for this analysis. It was developed using the MinkowskiEngine package~\cite{choy20194d} to handle sparse inputs more efficiently.}
\label{fig:cnn_architecture}
\end{figure*}

The main feature of CNNs is that they learn a series of filters (using convolutions), applied in sequence to extract increasingly powerful and abstract features that allow the CNN to learn a mapping between input images and target labels. Once the CNN is trained, it can be applied to new images to make accurate predictions on unseen examples during the training. The designed CNN architecture is depicted in Fig.~\ref{fig:cnn_architecture}. As introduced in Sec.~\ref{sec:intro}, it is based on 3D Submanifold Sparse Convolutions~\cite{graham2017submanifold} to deal with the sparse images used in this work. The CNN is trained\footnote{The rest of this paragraph applies to all the neural networks analysed in this article.} for 50 epochs\footnote{Epoch: one forward pass and one backward pass of all the training examples. In other words, an epoch is one pass over the entire dataset. For this paper, we show the test results for the epoch that performs best on the validation set.} using Python 3.6.9 and PyTorch 2.1.0~\cite{Abadi-et-al-2016-tensorflow}, as well as the Minkowski Engine package, version 0.5.4~\cite{choy20194d}, on an NVIDIA Tesla V100 GPUs. Stochastic Gradient Descent (SGD) is used as the optimiser, with a mini-batch size of 32 events, a learning rate of 0.1 (divided by 10 when the error plateaus, as suggested in~\cite{He-et-al-2015-deep}), a weight decay of 0.0001, and a momentum of 0.9\footnote{See Ref.~\cite{Goodfellow-et-al-2016-deep} for a description of optimisers and associated terminology.}. The model weights used for the analysis correspond to those at the epoch that maximises the overall accuracy on the validation set. There are no signs of overtraining.

\vspace{-0.25cm}

\subsection{Performance}
\label{sec:cnnresults}

\vspace{-0.25cm}

\begin{figure}[h]
\centering
\begin{center}
\includegraphics[width=0.925\linewidth, height = 5.5cm]{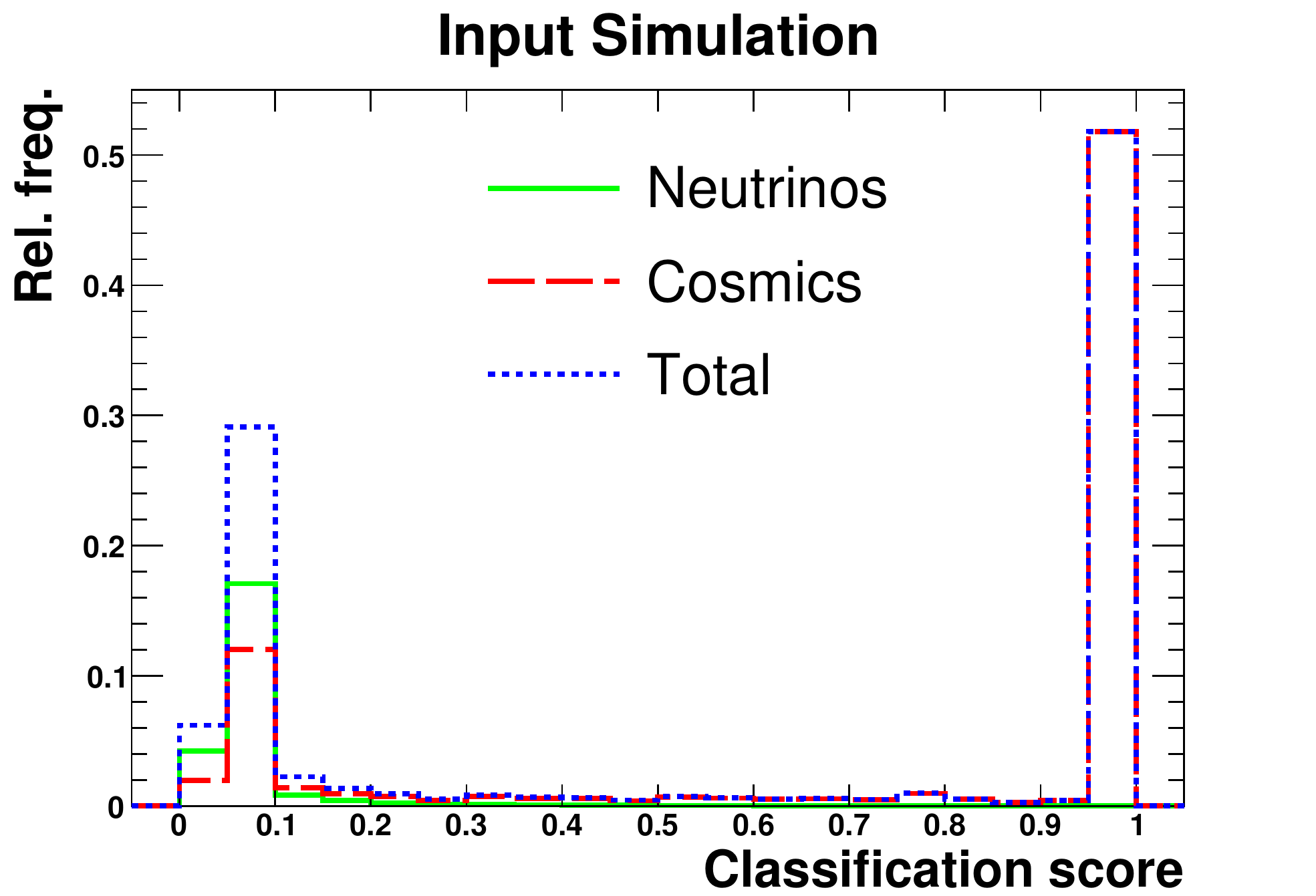}\\
\vspace{0.2cm}
\includegraphics[width=0.925\linewidth, height = 5.5cm]{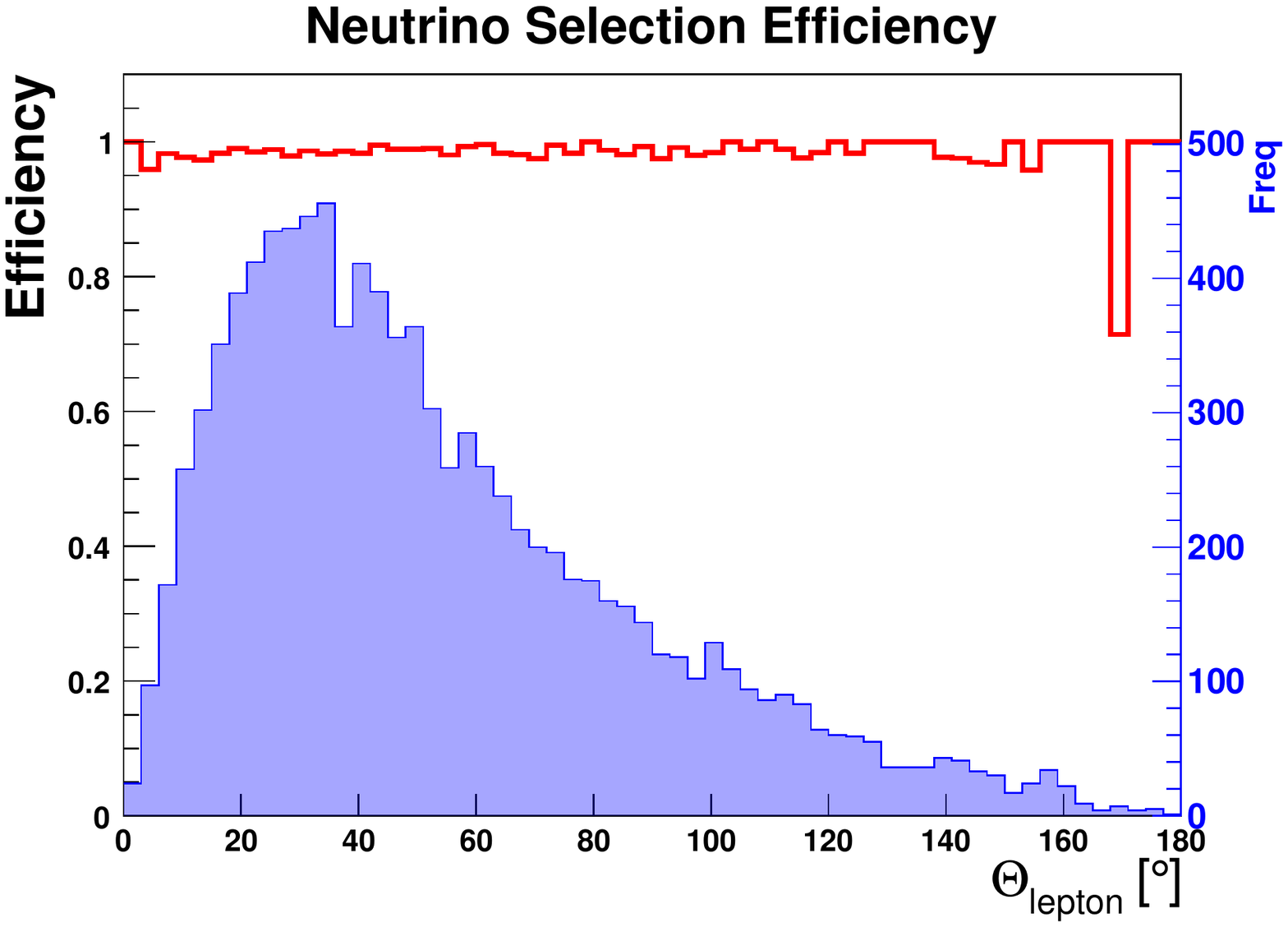}
\vspace{-0.1cm}
\caption{\label{fig:cnnscore}
The trained CNN's classification of events in the test sample (top). Note that the relative normalisation of cosmic and neutrino events is fixed to approximate what would be expected in data. The distribution of outgoing lepton angle (blue) with respect to the incoming neutrino from GENIE is shown alongside the neutrino selection efficiency of the CNN (red) following a cut at a classification score of 0.5 (bottom). }
\end{center}
\end{figure}

Once trained, the output of the CNN is a continuous score for each event between 0 (neutrino-like) and 1 (cosmic-like). The distribution of CNN scores for each true event type in the test sample is shown in Fig.~\ref{fig:cnnscore}. If a selection of neutrino events is made by cutting at a CNN score of 0.5, a 98.9\% selection efficiency is maintained whilst 76.3\% of cosmic-ray backgrounds are rejected. The charged-current selection efficiency was found to be flat (i.e., unbiased by kinematics) in various tested observables. An example of outgoing lepton angle is shown in Fig.~\ref{fig:cnnscore}.

\section{Reducing model-dependence with DANN-based training}
\label{sec:dannmethod}

\begin{figure*}[!htb]
\centering
\vspace{-0.2cm}
\includegraphics[width=13cm]{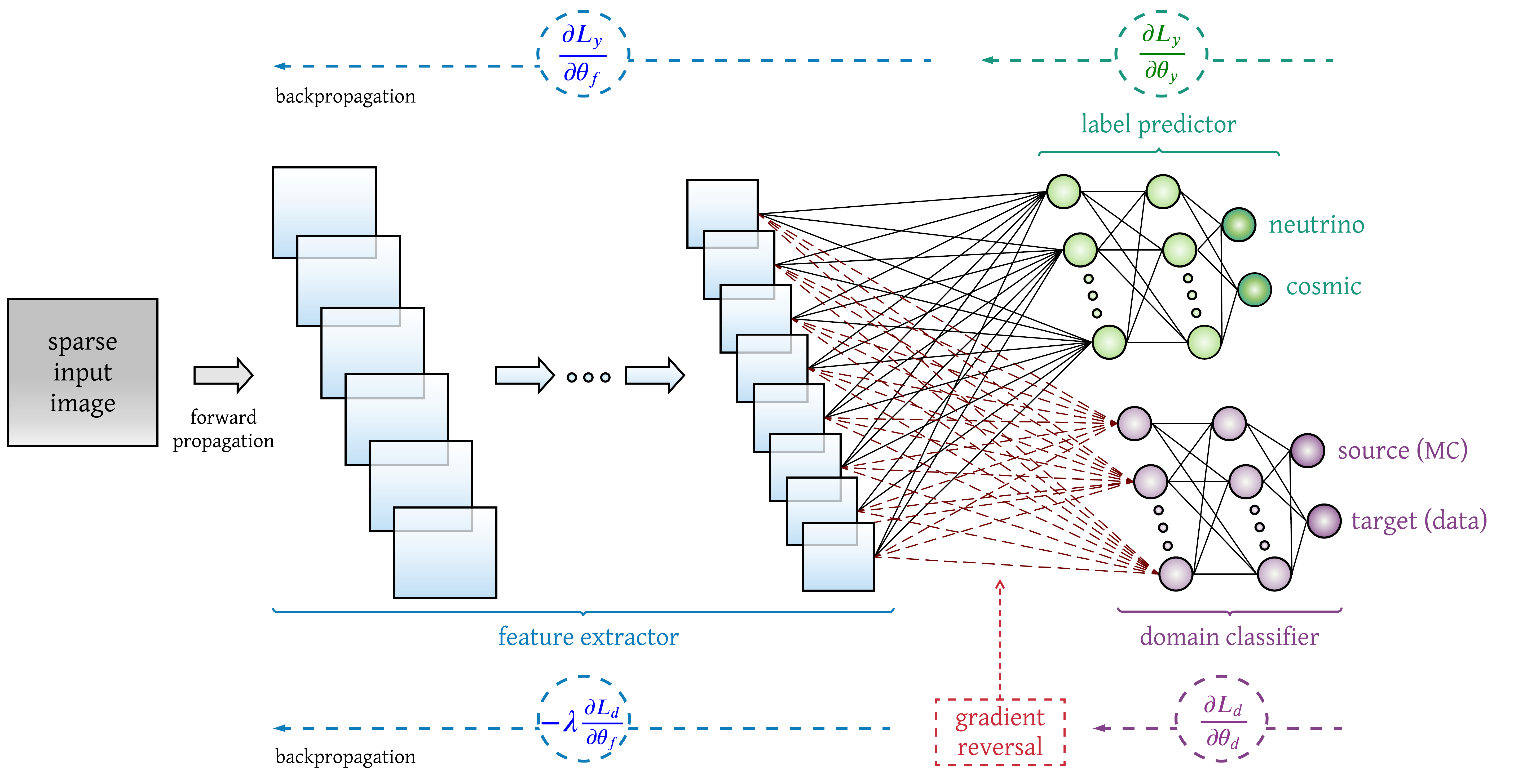}
\vspace{-0.1cm}
\caption{Domain-adversarial neural network architecture. The feature extractor (blue) and the label predictor (green) form the standard neural network classifier shown in Fig.~\ref{fig:cnn_architecture}. The domain classifier (purple) provides the domain adaptation part since it is connected to the feature extractor through a gradient reversal layer, allowing the alignment of feature distributions across the source and target domains. Figure adapted from~\cite{dann}.}
\label{fig:dann}
\end{figure*}

Whilst the CNN presented in Sec.~\ref{sec:cnnresults} shows excellent performance, the results assume perfect modelling of the neutrino and cosmic-ray events, the particle propagation and the detector response. If the CNN is trained with events that do not suitably represent what is in the real data, then the test sample's performance will not be reliable. Modern deep neural networks consist of millions (and sometimes billions) of parameters and have a strong representative capability with which they may exploit every detailed feature present in the simulation, including those that may not be present in data as well as others that may not follow the true physics model behind real data.
Thus, it is not easy to ever be sure that the pertinent aspects of the events are well modelled. To alleviate this issue, adversarial training methods can be employed to prevent neural networks from exploiting features that are only present in one of two domains. As a result, the performance can be made consistent in both domains. In this analysis, we show that it is possible to mitigate challenges associated with domain discrepancies through the application of DANNs.

In DANNs, the neural network model is trained on examples from two domains: (a) the \textit{source} domain, which consists of labelled simulated data; and (b) the \textit{target} domain, which consists of unlabelled true experimental data. The goal is to learn a discriminator from the labelled source domain examples and use the unlabelled target domain examples to ensure the discriminator relies on only domain-invariant features to perform the predictions. Regarding the implementation of the neural network, the classifier architecture remains identical, and it can be seen as the combination of a feature extractor (i.e., the bulk of the CNN, in our case) and a label predictor (i.e., the sparse linear layer(s) at the end). However, this alternative neural network has a second path, which connects the output of the feature extractor through a gradient reversal layer with a few linear layers that form a domain classifier. The gradient reversal layer performs an identity transformation during the forward propagation process but multiplies the gradient by a negative constant during the back-propagation, guarantying that the parameters learnt by the feature extractor are made similar for the source and target distributions. In other words, with this approach, the features learnt by this model are simultaneously discriminative - thanks to the label predictor - and domain-invariant - thanks to the domain classifier. This behaviour is shown in Fig.~\ref{fig:dann}. Furthermore, if some events from the target distribution are labelled (e.g., experimental data cosmic rays produced without a neutrino beam), those events might be used for the feature extractor learning too, making the domain adaptation semi-supervised, in contrast to the unsupervised case where all the events from the target distribution are unlabelled.

In order to test the effectiveness of DANNs as a method of reducing simulation dependence, we perform a series of mock-data studies. For these studies, statistically independent simulations of events (from neutrinos and cosmic rays) are produced before being modified to simulate possible mis-modelling bias. Since the coarse PMT information used in this analysis is likely not sensitive to the exact details of the neutrino interaction or cosmic-ray production, we focus primarily on applying distortions to the simulated detector response. The details of the mock data are as follows:

\begin{figure*}[htb]
\centering
\includegraphics[width=0.45\linewidth]{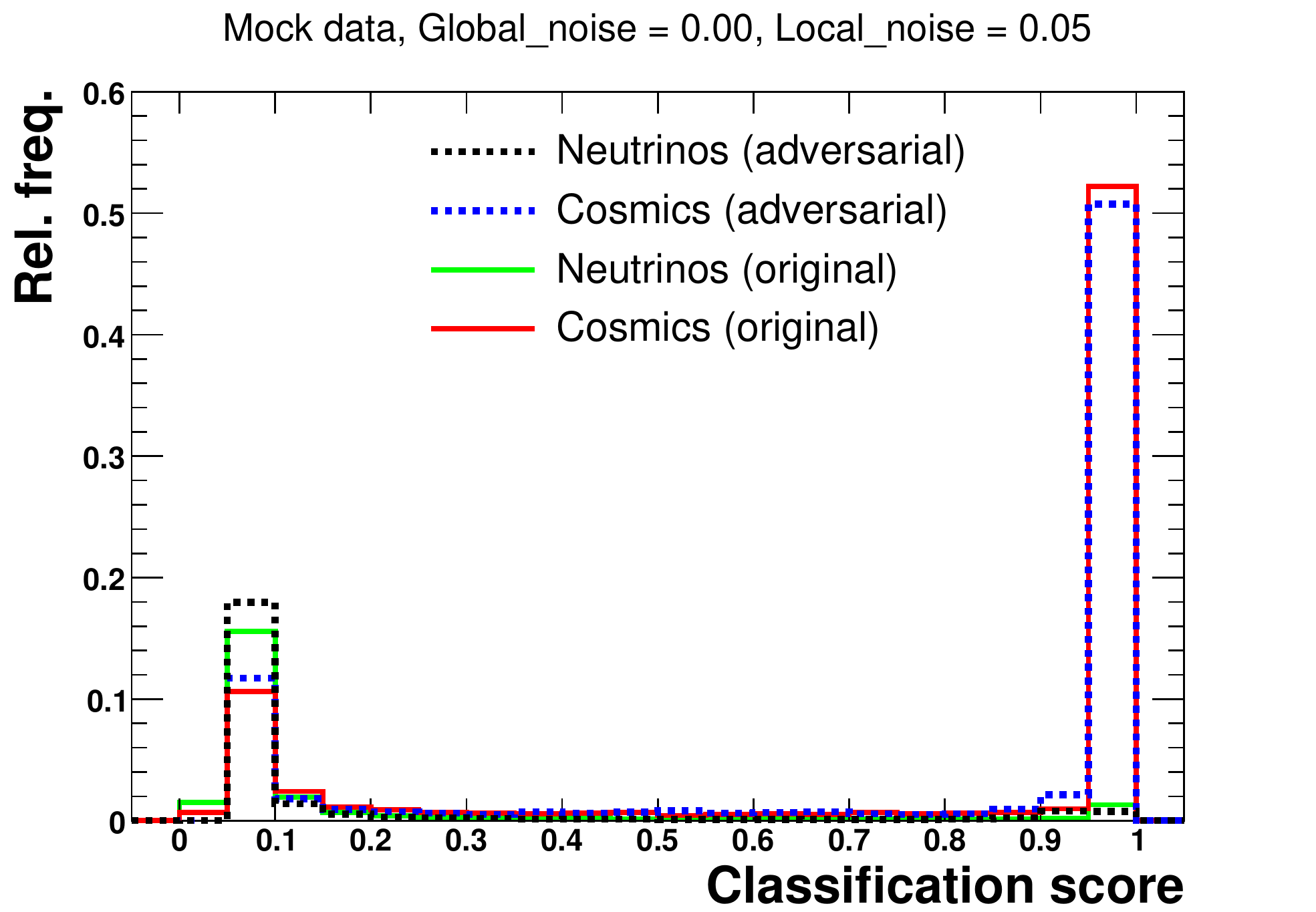}\hspace{0.3cm}
\includegraphics[width=0.45\linewidth]{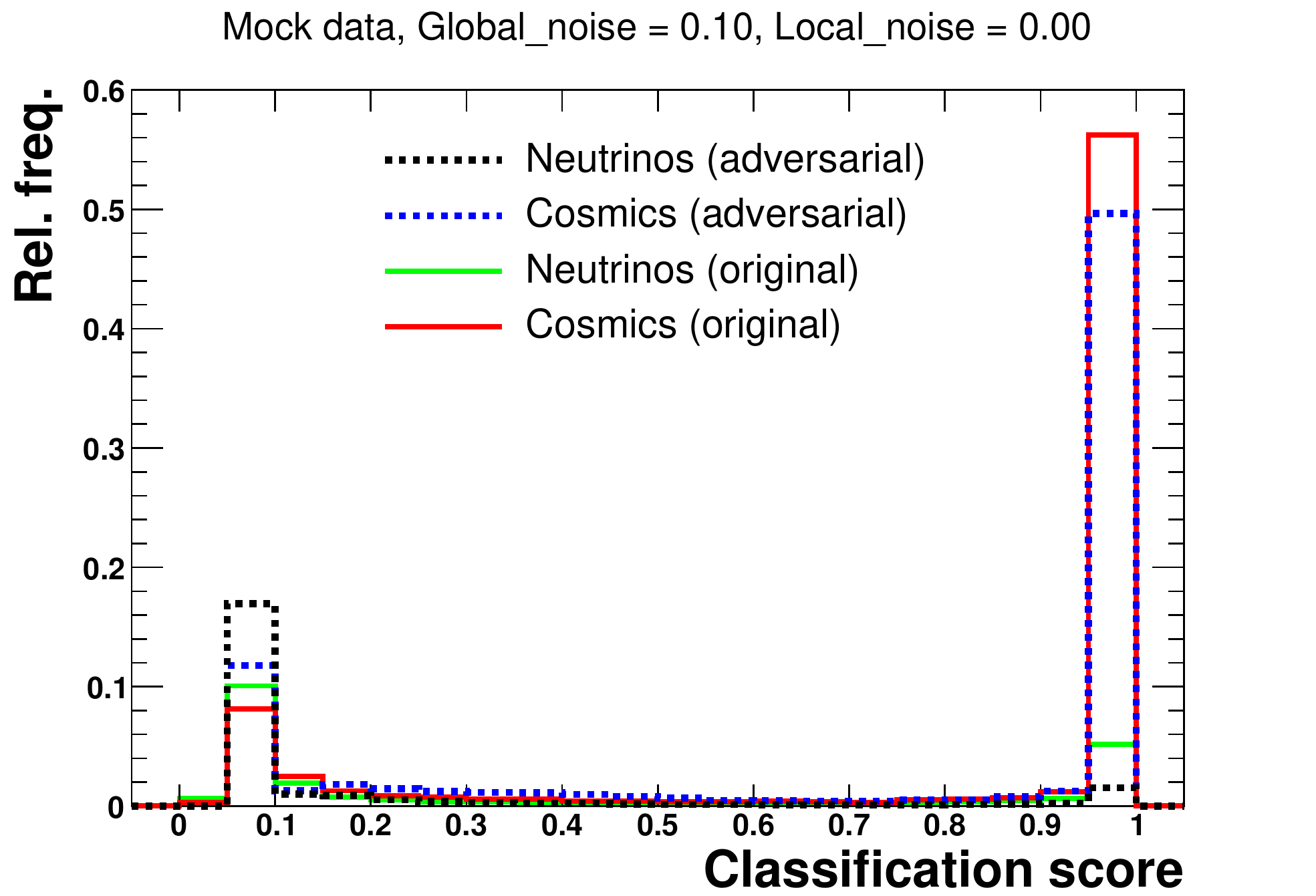}
\caption{\label{fig:dannscore} Results of the classification score of the nominal and adversarially trained CNN/DANN model applied to the original and: local noise mock data using a 5\% noise spread (left) and the global noise mock data using a 10\% noise spread (right). None of the cosmic-ray-mock-data events are labelled by event type.}
\end{figure*}

\paragraph*{``Global noise'' data:} in this mock data, noise, which is uncorrelated with the event content, is randomly added to each PMT with some pre-specified `global' probability that is common to all PMTs. The global noise probabilities considered are 2\%, 5\% and 10\%. The timing of the noise is considered as uniform distribution, and if noise is simulated to arrive before a PMT is opened by a simulated signal, the opening time of the signal is overwritten by that of the noise.

\paragraph*{``Local noise'' data:} similarly to the global noise data, this mock data set considers the addition of random noise to each PMT but where the probability of producing noise is different for every PMT. Noise probabilities per event for each PMT were generated randomly using a uniform distribution between 0 and either 2\%, 5\% or 10\%.

For each mock-data study, the DANN is trained as described in Sec~\ref{sec:cnnmethod} but with an addition of 22,778 cosmic-ray-mock-data events and 94,306 neutrino-mock-data events (40\% train, 10\% validation, 50\% test\footnote{Following the suggestion in \cite{perdue2018reducing}, we do not use all the available mock data events for training. Moreover, the large test set size provides enough statistics for the analysis.}), which are labelled by the domain (i.e., mock data or original simulation), all the simulated events are labelled by event type (i.e., cosmic or neutrino), and we consider three scenarios for the event type of mock-data events: (1) none of the events are labelled by event type (unsupervised domain adaptation), (2) 10\% of the cosmic-ray-mock-data events are labelled (semi-supervised domain adaptation), and (3) 50\% of the cosmic-ray-mock-data events are labelled (semi-supervised domain adaptation). This method could equally be applied to real data instead of mock data, using run periods with no neutrino beam to label the cosmic-ray events.

Both the originally trained CNN (as described in Sec.~\ref{sec:cnnmethod}) and the newly trained DANNs are used to attempt to classify events from the original sample and from the new mock data sample. An example of the classification scores for each model applied to the original and mock data sets is shown for two mock-data studies in Fig.~\ref{fig:dannscore}. A summary of the neutrino selection efficiency and the background rejection performance for the nominal simulation as well as for each mock data set is shown in Tab.~\ref{tab:mockdataperf}. The presented numbers are provided for a selection cut set to 0.5 of the network classification score in all cases. It should be noted that there is additional freedom to optimise performance as desired by tuning the cut value applied. For example, it was found that for the case of mock data with global noise of 10\%, the results using an unsupervised domain adaptation improve with respect to the original model by $\sim$14\% for neutrino selection efficiency and $\sim$1\% for cosmic-background rejection by setting a cut at 0.25. However, with the cut at 0.5, the neutrino selection efficiency improves dramatically (by $\sim$22\%), but the background rejection performance decreases (by $\sim$8\%). An alternative assessment of the CNN and DANN performance where the cut is varied to keep the background rejection factor constant is presented in Appendix~\ref{app:compresults}.

\begin{table*}[htb]
\vspace{0.5cm}
\small
    \centering
    \resizebox{0.8\linewidth}{!}{
    \begin{tabular}{l@{\hspace{0.25cm}}l@{\hspace{0.25cm}}rc@{\hspace{0.5cm}}rl@{\hspace{0.75cm}}rl@{\hspace{0.0cm}}rl@{\hspace{0.0cm}}rl}
    \hline \hline
    \multicolumn{3}{c}{\multirow{3}{*}{Dataset}} & \multicolumn{1}{l}{\multirow{3}{*}{Model type}} & \multicolumn{1}{c}{\multirow{3}{*}{\textbf{$\mathcal{E}_{\nu}$}[\%]}} & \multicolumn{1}{l}{\multirow{3}{*}{\textbf{$\mathcal{R}_{cos}$}[\%]}} &  \multicolumn{6}{c}{Domain adaptation} \\
     & & &   & 
    \multicolumn{2}{c}{ } & \multicolumn{2}{c}{Unsupervised} & \multicolumn{2}{c}{Semisup. 10\%} & \multicolumn{2}{c}{Semisup. 50\%}\\
     & & & & 
      &  & \multicolumn{1}{c}{\textbf{$\mathcal{E}_{\nu}$}[\%]} & \multicolumn{1}{c}{\textbf{$\mathcal{R}_{cos}$}[\%]} & \multicolumn{1}{c}{\textbf{$\mathcal{E}_{\nu}$}[\%]} & \multicolumn{1}{c}{\textbf{$\mathcal{R}_{cos}$}[\%]} & \multicolumn{1}{c}{\textbf{$\mathcal{E}_{\nu}$}[\%]} & \multicolumn{1}{c}{\textbf{$\mathcal{R}_{cos}$}[\%]} \\
    \hline
    \multicolumn{3}{c}{\multirow{7}{*}{Nominal}} & CNN & 98.9 & 76.3 & \multicolumn{2}{c}{-} & \multicolumn{2}{c}{-} & \multicolumn{2}{c}{-}\\
    \cline{4-12}
    &                                &  & DANNG2 & \multicolumn{2}{c}{-}  & 97.2 & 77.2 & 99.0 & 74.8 & 97.9 & 77.2\\ 
    &                                &  & DANNG5 & \multicolumn{2}{c}{-}  & 98.3 & 76.8 & 98.0 & 76.6 & 97.9 & 76.9\\ 
    &                                &  & DANNG10 & \multicolumn{2}{c}{-}  & 98.7 & 75.8 & 98.1 & 76.4 & 98.5 & 76.4\\

    \cline{4-12}
    &                                &  & DANNL2 & \multicolumn{2}{c}{-}  & 98.2 & 76.1 & 98.1 & 75.8 & 98.7 & 76.1\\ 
    &                                &  & DANNL5 & \multicolumn{2}{c}{-} & 98.8 & 75.9 & 98.7 & 76.0 & 98.8 & 76.2\\ 
    &                                &  & DANNL10 & \multicolumn{2}{c}{-} &  98.7 & 75.2 & 98.1 & 76.8 & 98.0 & 76.8\\ 
    
    \hline
    
    \multirow{12}{*}{\shortstack[l]{Mock\\data}}  & \multirow{6}{*}{\shortstack[l]{Global\\noise}}                                          & \multirow{2}{*}{2\%} & CNN & 91.7 & 74.8 & \multicolumn{2}{c}{-} & \multicolumn{2}{c}{-} & \multicolumn{2}{c}{-} \\ 
    &                               &                      & DANNG2 & \multicolumn{2}{c}{-} & 92.7 & 76.5 & 92.7 & 75.1 & 87.2 & 78.7 \\
    &                               & \multirow{2}{*}{5\%} & CNN & 81.0 & 75.8 & \multicolumn{2}{c}{-} & \multicolumn{2}{c}{-} & \multicolumn{2}{c}{-} \\
    &                               &                      & DANNG5 & \multicolumn{2}{c}{-} & 89.7 & 72.1 & 89.9  & 76.6 & 84.1 & 78.7 \\
    &                               & \multirow{2}{*}{10\%}& CNN & 66.4 & 79.0 & \multicolumn{2}{c}{-} & \multicolumn{2}{c}{-} & \multicolumn{2}{c}{-} \\
    &                               &                      & DANNG10 & \multicolumn{2}{c}{-} & 88.8 & 71.0 & 87.2 & 69.8 & 88.1 & 75.9 \\
                                   
    \cline{2-12}
    & \multirow{6}{*}{\shortstack[l]{Local\\noise}} & \multirow{2}{*}{2\%} & CNN & 95.5 & 75.1 & \multicolumn{2}{c}{-} & \multicolumn{2}{c}{-} & \multicolumn{2}{c}{-}\\ 
    &                               &                      & DANNL2 & \multicolumn{2}{c}{-} & 97.6 & 74.2 & 98.2 & 74.8 & 96.5 & 76.6 \\
    &                               & \multirow{2}{*}{5\%} & CNN & 90.2 & 74.8 & \multicolumn{2}{c}{-} & \multicolumn{2}{c}{-} & \multicolumn{2}{c}{-} \\
    &                               &                      & DANNL5 & \multicolumn{2}{c}{-} & 89.9 & 75.0 & 90.2 & 75.9 & 90.3 & 78.9 \\
    &                               & \multirow{2}{*}{10\%}& CNN & 81.9 & 75.7 & \multicolumn{2}{c}{-} & \multicolumn{2}{c}{-} & \multicolumn{2}{c}{-} \\
    &                               &                      & DANNL10 & \multicolumn{2}{c}{-} & 90.2 & 73.3 & 88.7 & 77.9 & 88.4 & 78.9 \\
    \hline \hline
    \end{tabular}}
\caption{\label{tab:mockdataperf}Efficiency (\textbf{$\mathcal{E}_{\nu}$}) and proportion of rejected cosmic-ray background events ($\mathcal{R}_{cos}$) using the original and adversarially trained CNN/DANN to classify events in the nominal simulations and mock-data studies. Table~\ref{tab:legend} details the model-type nomenclature.}
\end{table*}

\begin{table*}[htb]
\vspace{0.5cm}
\small
    \centering
    \resizebox{0.8\linewidth}{!}{
    \begin{tabular}{l@{\hspace{0.25cm}}l}
    \hline \hline
    Network name & Description \\
    \hline
    CNN & Original neural network trained on the nominal simulation.\\
    DANNG2 & Adversarial network trained on nominal simulation + mock data (global noise = 2\%).\\
    DANNG5 & Adversarial network trained on nominal simulation + mock data (global noise = 5\%).\\
    DANNG10 & Adversarial network trained on nominal simulation + mock data (global noise = 10\%).\\
    DANNL2 & Adversarial network trained on nominal simulation + mock data (local noise = 2\%).\\
    DANNL5 & Adversarial network trained on nominal simulation + mock data (local noise = 5\%).\\
    DANNL10 & Adversarial network trained on nominal simulation + mock data (local noise = 10\%).\\
    \hline \hline
    \end{tabular}}
\caption{\label{tab:legend}Networks legend.}
\end{table*}

These results show that, without the adversarial training, the original CNN can reject a sizeable portion of neutrino interactions in the mock data. However, once the adversarial training is used, the network is able to mitigate the bias and maintain a very high neutrino selection efficiency (the main goal of the filter) whilst continuing to achieve a significant rejection of cosmic-ray backgrounds. This occurs for the unsupervised and the semi-supervised domain adaptations, and it is more visible for larger noises (i.e., 10\% of global noise and 10\% of local noise). It can equally be observed that the use of a DANN does not degrade the performance of the baseline CNN when applied to the nominal simulation. This demonstrates that even if the simulation very well describes the data, the use of a DANN over a CNN is not expected to degrade the performance.

Concerning the unsupervised domain adaptation compared to the semi-supervised cases, we find a small but non-negligible improvement in the rejection of mock-data cosmic-ray backgrounds for some mock data studies, while suffering a slight reduction in mock-data neutrino selection efficiency. This behaviour is expected since, for the semi-supervised cases, the models have more labelled cosmic events to learn to reject from (especially for the case where 50\% of the mock-data cosmic rays are labelled). It is possible that labelling a larger portion of the training sample may allow improved performance from the semi-supervision (recall that the majority of events in the training sample are neutrino interactions, which cannot be labelled in real data).

\section{Conclusion}
\label{sec:discussion}

The studies presented in this manuscript demonstrate that easily accessible information from LAr-TPC experiment's light detection systems, which requires very little processing, may be used to effectively separate neutrino from cosmic-ray induced signals within a neutrino beam spill. The use of a specially adapted CNN ensures that the majority of cosmic-ray interactions can be filtered out without the rejection of almost any neutrino induced interactions. 

Whilst the use of a CNN trained on simulated event samples is susceptible to bias due to mis-modelling, potentially causing the inadvertent rejection of neutrino events, it is demonstrated that adversarial training via a DANN can mitigate the loss of efficiency at the cost of some reduced background rejection. It is further shown that in some cases, the background rejection performance may be improved through semi-supervised domain adaptation of the DANN using labelled real cosmic ray events. 

Overall the techniques presented in this manuscript demonstrate a method for providing a significant rejection of cosmic-ray events without the need for computationally expensive reconstruction algorithms. These methods are shown to be effective when applied to simulations from the ICARUS experiment, but are easily adaptable and could likely achieve similar success if applied to other LAr-TPC experiments. 

\bibliographystyle{unsrt}
\bibliography{biblio}

\begin{thebibliography}{10}

\bibitem{Rubbia:1977zz}
C.~Rubbia.
\newblock {The Liquid Argon Time Projection Chamber: A New Concept for Neutrino
  Detectors}.
\newblock 5 1977.

\bibitem{Willis:1974gi}
W.~J. Willis and V.~Radeka.
\newblock {Liquid Argon Ionization Chambers as Total Absorption Detectors}.
\newblock {\em Nucl. Instrum. Meth.}, 120:221--236, 1974.

\bibitem{ArgoneuT:2016wjb}
R.~Acciarri et~al.
\newblock {First Observation of Low Energy Electron Neutrinos in a Liquid Argon
  Time Projection Chamber}.
\newblock {\em Phys. Rev. D}, 95(7):072005, 2017.

\bibitem{Antonello:2015lea}
M.~Antonello et~al.
\newblock {A Proposal for a Three Detector Short-Baseline Neutrino Oscillation
  Program in the Fermilab Booster Neutrino Beam}.
\newblock 3 2015.

\bibitem{Abi:2020wmh}
B.~Abi et~al.
\newblock {Deep Underground Neutrino Experiment (DUNE), Far Detector Technical
  Design Report, Volume I Introduction to DUNE}.
\newblock {\em JINST}, 15(08):T08008, 2020.

\bibitem{firstCNN}
Y.~LeCun et~al.
\newblock Backpropagation applied to handwritten zip code recognition.
\newblock {\em Neural Computation}, 1(4):541--551, Dec 1989.

\bibitem{LeCun-et-al-1998-gradient}
Y.~LeCun, L.~Bottou, Y.~Bengio, and P.~Haffner.
\newblock Gradient-based learning applied to document recognition.
\newblock {\em Proceedings of the IEEE}, 86(11):2278--2324, 1998.

\bibitem{Aurisano:2016jvx}
A.~Aurisano et~al.
\newblock {A Convolutional Neural Network Neutrino Event Classifier}.
\newblock {\em JINST}, 11(09):P09001, 2016.

\bibitem{DUNE:2020gpm}
B.~Abi et~al.
\newblock {Neutrino interaction classification with a convolutional neural
  network in the DUNE far detector}.
\newblock {\em Phys. Rev. D}, 102(9):092003, 2020.

\bibitem{MicroBooNE:2018kka}
C.~Adams et~al.
\newblock {Deep neural network for pixel-level electromagnetic particle
  identification in the MicroBooNE liquid argon time projection chamber}.
\newblock {\em Phys. Rev. D}, 99(9):092001, 2019.

\bibitem{graham2017submanifold}
B.~Graham and L.~van~der Maaten.
\newblock Submanifold sparse convolutional networks.
\newblock 2017.

\bibitem{Adams-2019-deep}
C.~Adams et~al.
\newblock Deep neural network for pixel-level electromagnetic particle
  identification in the {MicroBooNE} liquid argon time projection chamber.
\newblock {\em Phys. Rev. D}, 99:092001, May 2019.

\bibitem{Domine-2020-scalable}
L.~Domin\'e and K.~Terao.
\newblock Scalable deep convolutional neural networks for sparse, locally dense
  liquid argon time projection chamber data.
\newblock {\em Phys. Rev. D}, 102:012005, Jul 2020.

\bibitem{SperdutiFirstGNN}
A.~{Sperduti} and A.~{Starita}.
\newblock Supervised neural networks for the classification of structures.
\newblock {\em IEEE Transactions on Neural Networks}, 8(3):714--735, 1997.

\bibitem{zhou2018graph}
Jie Zhou, Ganqu Cui, Zhengyan Zhang, Cheng Yang, Zhiyuan Liu, Lifeng Wang,
  Changcheng Li, and Maosong Sun.
\newblock Graph neural networks: A review of methods and applications, 2018.

\bibitem{ben2010theory}
S.~Ben-David et~al.
\newblock A theory of learning from different domains.
\newblock {\em Machine learning}, 79(1):151--175, 2010.

\bibitem{redko2019advances}
I.~Redko et~al.
\newblock {\em Advances in domain adaptation theory}.
\newblock Elsevier, 2019.

\bibitem{dann}
Y.~Ganin et~al.
\newblock Domain-adversarial training of neural networks.
\newblock {\em J. Mach. Learn. Res.}, 17(1):2096–2030, January 2016.

\bibitem{perdue2018reducing}
GN~Perdue, A~Ghosh, M~Wospakrik, F~Akbar, DA~Andrade, M~Ascencio, L~Bellantoni,
  A~Bercellie, M~Betancourt, GFR~Caceres Vera, et~al.
\newblock Reducing model bias in a deep learning classifier using domain
  adversarial neural networks in the minerva experiment.
\newblock {\em Journal of Instrumentation}, 13(11):P11020, 2018.

\bibitem{Amerio:2004ze}
S.~Amerio et~al.
\newblock {Design, construction and tests of the ICARUS T600 detector}.
\newblock {\em Nucl. Instrum. Meth. A}, 527:329--410, 2004.

\bibitem{Babicz:2018svg}
M.~Babicz et~al.
\newblock {Test and characterization of 400 Hamamatsu R5912-MOD photomultiplier
  tubes for the ICARUS T600 detector}.
\newblock {\em JINST}, 13(10):P10030, 2018.

\bibitem{Ali-Mohammadzadeh:2020fbd}
B.~Ali-Mohammadzadeh et~al.
\newblock {Design and implementation of the new scintillation light detection
  system of ICARUS T600}.
\newblock {\em JINST}, 15(10):T10007, 2020.

\bibitem{Babicz:2020rpf}
M.~Babicz et~al.
\newblock {A particle detector that exploits Liquid Argon scintillation light}.
\newblock {\em Nucl. Instrum. Meth. A}, 958:162421, 2020.

\bibitem{Heck:1998vt}
D.~Heck, J.~Knapp, J.~N. Capdevielle, G.~Schatz, and T.~Thouw.
\newblock {CORSIKA: A Monte Carlo code to simulate extensive air showers}.
\newblock 2 1998.

\bibitem{Agostinelli:2002hh}
S.~Agostinelli et~al.
\newblock {GEANT4--a simulation toolkit}.
\newblock {\em Nucl. Instrum. Meth. A}, 506:250--303, 2003.

\bibitem{Snider:2017wjd}
E.~L. Snider and G.~Petrillo.
\newblock {LArSoft: Toolkit for Simulation, Reconstruction and Analysis of
  Liquid Argon TPC Neutrino Detectors}.
\newblock {\em J. Phys. Conf. Ser.}, 898(4):042057, 2017.

\bibitem{AguilarArevalo:2008yp}
A.~A. Aguilar-Arevalo et~al.
\newblock {The Neutrino Flux prediction at MiniBooNE}.
\newblock {\em Phys. Rev. D}, 79:072002, 2009.

\bibitem{Andreopoulos:2009rq}
C.~Andreopoulos et~al.
\newblock {The GENIE Neutrino Monte Carlo Generator}.
\newblock {\em Nucl. Instrum. Meth. A}, 614:87--104, 2010.

\bibitem{choy20194d}
C.~Choy, J.~Gwak, and S.~Savarese.
\newblock 4d spatio-temporal convnets: Minkowski convolutional neural networks.
\newblock 2019.

\bibitem{Abadi-et-al-2016-tensorflow}
M.~Abadi et~al.
\newblock Tensorflow: A system for large-scale machine learning.
\newblock In {\em OSDI}, volume~16, pages 265--283, 2016.

\bibitem{He-et-al-2015-deep}
K.~He, X.~Zhang, S.~Ren, and J.~Sun.
\newblock Deep residual learning for image recognition.
\newblock {\em CoRR}, 2015.

\bibitem{Goodfellow-et-al-2016-deep}
I.~Goodfellow, Y.~Bengio, and A.~Courville.
\newblock {\em Deep Learning}.
\newblock MIT Press, 2016.
\newblock \url{http://www.deeplearningbook.org}.

\end{thebibliography}

\section*{Acknowledgements}

The authors would like to thank the ICARUS collaboration for supporting this work. Particular thanks is given to François Drieslma for providing detailed comments on a draft version of this manuscript. The work of M. Babicz was supported by the National Science Center, Poland, research project No. 2019/33/N/ST2/02874.\\

\appendix

\section{Complementary results}
\label{app:compresults}

\begin{table*}[htb]
\small
    \centering
    \resizebox{0.75\linewidth}{!}{
    \begin{tabular}{l@{\hspace{0.25cm}}l@{\hspace{0.25cm}}rc@{\hspace{.2cm}}c@{\hspace{1.5cm}}c@{\hspace{0.0cm}}c@{\hspace{0.0cm}}c}
    \hline \hline
    
    \multicolumn{3}{c}{\multirow{3}{*}{Dataset}} & \multicolumn{1}{l}{\multirow{3}{*}{Model type}} & \multicolumn{1}{c}{\multirow{3}{*}{\textbf{$\mathcal{E}_{\nu}$}[\%]}} &  \multicolumn{3}{c}{Domain adaptation} \\
     & & &   & 
    \multicolumn{1}{c}{ } & \multicolumn{1}{c}{Unsupervised} & \multicolumn{1}{c}{Semisup. 10\%} & \multicolumn{1}{c}{Semisup. 50\%}\\
     & & & & 
      & \multicolumn{1}{c}{\textbf{$\mathcal{E}_{\nu}$}[\%]} &  \multicolumn{1}{c}{\textbf{$\mathcal{E}_{\nu}$}[\%]} & \multicolumn{1}{c}{\textbf{$\mathcal{E}_{\nu}$}[\%]} \\
    \hline
    \multicolumn{3}{c}{\multirow{7}{*}{Nominal}} & CNN & \multicolumn{1}{c}{99.2}  & \multicolumn{1}{c}{-} & \multicolumn{1}{c}{-} & \multicolumn{1}{c}{-}\\
    \cline{4-8}
    &                                &  & DANNG2 & \multicolumn{1}{c}{-}  & 98.3 & 98.9 & 98.8\\ 
    &                                &  & DANNG5 & \multicolumn{1}{c}{-}  & 99.0 & 98.7 & 98.9\\
    &                                &  & DANNG10 & \multicolumn{1}{c}{-}  & 99.0 & 98.6 & 98.7\\

    \cline{4-8}
    &                                &  & DANNL2 & \multicolumn{1}{c}{-}  & 98.6 & 98.3 & 99.1\\ 
    &                                &  & DANNL5 & \multicolumn{1}{c}{-} & 99.1 & 98.9 & 99.0\\ 
    &                                &  & DANNL10 & \multicolumn{1}{c}{-} & 98.7 & 98.8 & 98.8\\ 
    
    \hline
    
    \multirow{12}{*}{\shortstack[l]{Mock\\data}}  & \multirow{6}{*}{\shortstack[l]{Global\\noise}}                                          & \multirow{2}{*}{2\%} & CNN & \multicolumn{1}{c}{91.5} & \multicolumn{1}{c}{-} & \multicolumn{1}{c}{-} & \multicolumn{1}{c}{-} \\ 
    &                               &                      & DANNG2 & \multicolumn{1}{c}{-} & 93.6 & 92.7 & 89.6 \\
    &      & \multirow{2}{*}{5\%} & CNN & \multicolumn{1}{c}{81.7} & \multicolumn{1}{c}{-} & \multicolumn{1}{c}{-} & \multicolumn{1}{c}{-} \\
    &                               &                      & DANNG5 & \multicolumn{1}{c}{-} & 88.0 & 90.8 & 87.3 \\
    &      & \multirow{2}{*}{10\%}& CNN & \multicolumn{1}{c}{74.7} & \multicolumn{1}{c}{-} & \multicolumn{1}{c}{-} & \multicolumn{1}{c}{-} \\
    &                               &                      & DANNG10 & \multicolumn{1}{c}{-} & 86.0 & 84.5 & 88.6 \\
                                   
    \cline{2-8}
    & \multirow{6}{*}{\shortstack[l]{Local\\noise}} & \multirow{2}{*}{2\%} & CNN & \multicolumn{1}{c}{95.4} & \multicolumn{1}{c}{-} & \multicolumn{1}{c}{-} & \multicolumn{1}{c}{-}\\ 
    &                               &                      & DANNL2 & \multicolumn{1}{c}{-} & 97.1 & 97.9 & 97.1 \\
    &                               & \multirow{2}{*}{5\%} & CNN & \multicolumn{1}{c}{89.9} & \multicolumn{1}{c}{-} & \multicolumn{1}{c}{-} & \multicolumn{1}{c}{-} \\
    &                               &                      & DANNL5 & \multicolumn{1}{c}{-} & 89.9 & 90.6 & 93.7 \\
    &                               & \multirow{2}{*}{10\%}& CNN & \multicolumn{1}{c}{82.6} & \multicolumn{1}{c}{-} & \multicolumn{1}{c}{-} & \multicolumn{1}{c}{-} \\
    &                               &                      & DANNL10 & \multicolumn{1}{c}{-} & 89.1 & 90.5 & 91.3 \\
    \hline \hline
    \end{tabular}}
\caption{\label{tab:comresults}Efficiency (\textbf{$\mathcal{E}_{\nu}$}) using the original and adversarially trained CNN/DANN to classify events in the nominal simulations and mock-data studies. In contrast to what is shown in Table~\ref{tab:mockdataperf} and for a better comparison of the efficiencies, the classification score cuts were tuned so that the proportion of rejected cosmic-ray background events for each model is always 75\%. Table~\ref{tab:legend} details the model-type nomenclature.}
\end{table*}

As discussed in Sec.~\ref{sec:dannmethod}, the performance of the CNN and DANNs' application to the mock-data sets may be demonstrated in alternative ways to as presented in Tab.~\ref{tab:mockdataperf}, which shows the efficiency and background rejection achieved for a fixed cut in the networks' output classification score distribution. As such, Tab.~\ref{tab:comresults} instead changes the cut such that the background rejection remains fixed (at 75\%) so that the efficiencies can be more directly compared. The conclusions remain unchanged from those presented in Sec.~\ref{sec:dannmethod}: the performance improvement offered by the adversarial training of the DANNs is substantial with respect to naively applying CNN trained only on the input simulation. The improvement can be seen to be stronger for more extreme fake data studies. Labelling some proportion of the cosmic ray events in the mock-data samples to provide a semi-supervised of the DANNs can offer a small additional improvement in some cases.

\end{document}